\documentclass{pnastwo}

\usepackage[dvips]{graphicx}

\usepackage{amssymb,amsfonts,amsmath}

\contributor{Submitted to Proceedings
of the National Academy of Sciences of the United States of America}
\url{www.pnas.org/cgi/doi/10.1073/pnas.0709640104}
\copyrightyear{2008}
\issuedate{Issue Date}
\volume{Volume}
\issuenumber{Issue Number}

\begin{document}



\title{Indirect Detection of Dark Matter with $\gamma$ rays}





\author{Stefan Funk\affil{1}{Kavli Institute for Particle Astrophysics
  and Cosmology, Stanford University \& SLAC National Accelerator
  Center}}

\contributor{Submitted to Proceedings of the National Academy of Sciences
of the United States of America}

\maketitle

\begin{article}

  \begin{abstract} 
    The details of what constitutes the majority of the mass that
    makes up dark matter in the Universe remains one of the prime
    puzzles of cosmology and particle physics today - eighty years
    after the first observational indications. Today, it is widely
    accepted that dark matter exists and that it is very likely
    composed of elementary particles -- that are weakly interacting
    and massive (WIMPs for Weakly Interacting Massive
    Particles). As important as dark matter is in our understanding of
    cosmology, the detection of these particles has so far been
    elusive. Their primary properties such as mass and
    interaction cross sections are still unknown. Indirect
    detection searches for the products of
    WIMP annihilation or decay. This isq generally done through
    observations of gamma-ray photons or cosmic rays. Instruments such
    as the Fermi-LAT, H.E.S.S., MAGIC and VERITAS, combined with the
    future Cherenkov Telescope Array (CTA) will provide important
    complementarity to other search techniques. Given the expected
    sensitivities of all search techniques, we are at a stage where the
    WIMP scenario is facing stringent tests and it can be expected
    that WIMPs will be either be detected or the scenario will be so
    severely constrained that it will have to be re-thought. In this
    sense we are on the {\emph{``Threshold of Discovery''}}. In this
    article, I will give a general overview over the current status
    and the future expectations for indirect searches for dark matter
    (WIMP) particles.
 \end{abstract}

\keywords{Gamma rays | Dark Matter | Cosmology }



\dropcap{T}here is a broad consensus that dark matter is made up of
elementary particles. The most promising candidates are weakly
interacting massive particles (WIMPs), particularly if they also form
the lightest supersymmetric particle. The general assumption is that
the thermal freeze-out in the early Universe leaves a relic density of
dark matter particles in the current Universe (after the freeze-out
the particles become too diluted to annihilate in appreciable numbers
and thermal energies were too low to produce them. The co-moving
density is therefore roughly constant since then). The annihilation of
these particles into standard-model particles controls the abundance
in the Universe, there is thus a tight connection between the
annihilation cross section and cosmologically relevant quantities. 
For particles annihilating (in the simplest case, i.e.\ annihilating through
S waves~\cite{Griest})
the relic density only depends on the annihilation cross
section $\sigma_{\mathrm{ann}}$ weighted by the average velocity of
the particle (see e.g.\cite{JungmannKamionkowskiGriest1996}):

\begin{equation*}
\Omega_{\chi} h^2 = 0.11\, \frac{3\times
  10^{-26}\mathrm{cm}^3\mathrm{s}^{-1}}{<\sigma_{\mathrm{ann}}v>} 
\end{equation*}

As the value for the relic dark matter density from CMB observations
is $\Omega_{\chi}h^2 = 0.113 \pm 0.004$~\cite{Planck}, it follows that
the expected velocity-weighted annihilation cross-section is in the
range of $3\times 10^{-26}\mathrm{cm}^3\mathrm{s}^{-1}$. This
represents a striking connection that for typical gauge couplings to
ordinary standard model particles and a dark matter mass at the weak
interaction scale WIMPs have the right relic density (using standard
early Universe conditions) to match those of the cosmologically
measured dark matter particles. In other words, the value for the
interaction rate $<\sigma_{\mathrm{Ann}}v>$ corresponds to a cross
section of approximately 1 pb, i.e. a typical weak interaction cross
section.  This is the so-called WIMP miracle in which particles that
are motivated by a microphysical puzzle (or better a gauge hierarchy
problem) are excellent dark matter candidates. Obviously, this
connection could be merely a coincidence but if true, then naturally,
WIMP masses would be expected in the range of 10 GeV and a few TeV
which is why a lot of attention has been devoted to exploring that
mass range in the dark matter parameter space.

Given the tight connection between the amount of WIMP dark matter in
the current Universe and the annihilation cross-section it is natural
to expect dark matter self-annihilations. To be able to
self-annihilate the dark matter particle much either be a Majorana
particle or a Dirac particle with no matter-antimatter asymmetry. In
the annihilation (or decay) of the dark matter particles all kinds of
standard model particles are created (quarks, bosons, leptons) and
then produce either gamma rays or cosmic rays. In
particular regions in the Universe with high dark matter densities
(such as the centers of galaxies, and clusters of galaxies) have
enhanced probabilities that dark matter particles encounter each other
and annihilate. With an appropriate assumption about the density
distribution of dark matter (e.g.\ from numerical simulations) one can
predict the expected annihilation signal when assuming a certain
interaction rate $\sigma v$ or put limits on the latter in the absence
of a signal.

A more quantitative description of the expected flux of particles from
dark matter annihilation can be drawn from the following relation (for
more details, see e.g.\ the excellent review by~\cite{BergstroemUllioBuckley}):
\begin{equation*}
\frac{\mathrm{d} \Phi_\gamma}{\mathrm{d E}_\gamma} = \frac{1}{4\pi}
\underbrace{\frac{<\sigma_{\mathrm{ann}} v>}{2 \mathrm{m}^2_{\mathrm{WIMP}}}
\sum_f \frac{\mathrm{d N}_\gamma^f}{\mathrm{d E}_\gamma}
\mathrm{B}_f}_\text{'Particle Physics'}
\times
\underbrace{\int\limits_{\Delta \Omega} \mathrm{d}\Omega^{\prime}
\int\limits_{\mathrm{los}} \rho^2\mathrm{d}l(r,
\theta^{\prime})}_{\text{'Astrophysics' or} J(E)}
\end{equation*}

The left-hand side contains the (measureable) gamma ray flux. The
right-hand side contains two components (1) a particle physics term
which is given by the velocity-averaged annihilation cross-section
($<\sigma_{\mathrm{ann}} v>$), the mass of the dark matter particle
($m$) and the sum over the gamma-ray yields for a certain annihilation
channel into channel $f$ ($\mathrm{d N}_\gamma^f/\mathrm{d E}_\gamma$)
multiplied by the branching ratio into that channel ($\mathrm{B}_f$),
and (2) an astrophysics term $J(E)$ (called the J-factor) given by
the line-of-sight integral of the square of the dark matter density
$\rho$. Given that both the particle physics and the astrophysics
term are unknown, one needs to make an assumption about one in order to
put constraints on the other when measuring a gamma-ray flux (or an
upper limit). This in turn already points to one of the major
challenges in the indirect detection of dark matter: the astrophysical
uncertainties, both in the density profile of dark matter (which
enters quadratically) and in the suppression of the astrophysical
foregrounds (which affect the sensitivity or the minimal gamma-ray
flux that can be detected). In order to derive meaningful dark matter
limits, the astrophysical foregrounds have to be understood and
subtracted. For excellent general recent reviews on selected topics
related to the indirect detection of dark matter, see~\cite{Cirelli,
  Profumo, Lavalle, Bringmann}. 

While most recent studies to detect the secondary products of dark
matter annihilations have focussed on gamma rays, annihilations into
cosmic rays can also be used. Given the large flux of cosmic rays
accelerated directly in astrophysical sources (primary cosmic rays)
and produced in the interaction of cosmic rays with interstellar
material (secondary cosmic rays) it is beneficial to use particles
that are less-frequently produced in these settings. The most commonly
used are antimatter particles, in particular anti-deuterons,
anti-protons and positrons, which are not so copiously produced in
astrophysics accelerators. These can provide important clues towards
the dark matter puzzle as e.g.\ seen in the rise in the positron
fraction recently observed by PAMELA~\cite{PositronPAMELA} and
confirmed by the Fermi-LAT and AMS~\cite{PositronFermi,
  PositronAMS}. However, I will mention them in this article only in
passing and will focus on gamma-ray observations.

Gamma rays can be produced by dark matter annihilations in two major
ways: (a) continuum signals from annihilation into other particles
which eventually produces gamma rays either through pion production,
or final state bremsstrahlung and inverse Compton from leptonic
channels and (b) line signals from dark matter annihilating directly
to $\gamma$X, where X usually is another neutral state, typically
$\gamma$ ray or Z or a Higgs boson. Given that dark matter particles
are essentially at rest (for cold dark matter), the photons will
emerge back-to-back with an energy directly related to the rest mass
of the dark matter particle $E_\gamma = m_\chi$ or $E_\gamma =
m_\chi(1-m_X/4 M_\chi^2)$. While the line signal can provide a
``smoking gun'' signal for dark matter annihilation, its flux is
typically loop suppressed by a factor of $\alpha_e^2$ where $\alpha_e$
is the fine structure constant (the electrically neutral dark matter
particle does not couple directly to photons but has to go through a
charged particle loop) . The signal is therefore expected to be much
smaller than the continuum flux. This continuum signal has a smooth
energy distribution with an exponential cutoff at the mass of the dark
matter particle $E_{\gamma} = m_{\chi}$. The spectral shape is
universal in the sense that it takes a similar form for almost any
channel and depends somewhat weakly on $m_{\chi}$. The exact
annihilation channel depends on the properties of the WIMP but is
typically (for bino-like WIMPs) dominated by annihilation into
$b\bar{b}$ pairs with pair production into $\tau$-leptons also
contributing. For more massive WIMPs with a wino or higgsino component
annihilation will proceed into massive gauge bosons. Annihilations
that have a large branching fraction into $e^{+}e^{-}$ pairs will
enhance the gamma-ray signal through inverse Compton scattering these
on starlight or the CMB (see e.g.~\cite{Baltz, Regis}).

Gamma rays have several unique properties which make them ideally
suited to study dark matter annihilations. Primarily, they do not get
deflected in intervening magnetic fields and thus point back at the
site at which they are created. This allows one to search for gamma
ray signatures not only in our vicinity in the Galaxy but also in
distant objects such as satellite galaxies or even galaxy clusters. In
the case of a signal, this could make provide a unique method to study
the distribution of dark matter in our Galaxy or in the Universe. The
energy of gamma rays from dark matter annihilation is limited by the
rest mass of the annihilating particle. Gamma rays therefore also
provide a unique spectral signature. In particular, given the
aforementioned preferred WIMP masses at the scale of a weakly
interacting particle, gamma-rays in the GeV and TeV range access the
most relevant mass range of dark matter particles. One final advantage
of the usage of gamma rays is that, in the local Universe gamma-rays do
not suffer from attenuation, and therefore they retain the source
spectral information intact at the Earth.

In the following section~\ref{sec::results} I will summarize current
observational results with a focus on gamma-ray
observations. Currently, the most productive observatory for
dark-matter related publications is the Fermi-LAT with currently
$\sim$200 such refereed publications. The Fermi-LAT has recently
provided a major breakthrough in the indirect detection of dark matter
by reaching for the first time below the aforementioned thermal relic
annihilation cross-section of $3\times 10^{-26}
\mathrm{cm}^3\mathrm{s}^{-1}$ through observations of dwarfs
spheroidals. The final section will give an outlook for future
prospects in indirect dark matter observations and will also discuss
the complementarity of the various search methods.

\section{Observational results}
\label{sec::results}

\begin{figure*}[t]
\begin{center}
\resizebox{1.8
  \columnwidth}{!}{\includegraphics{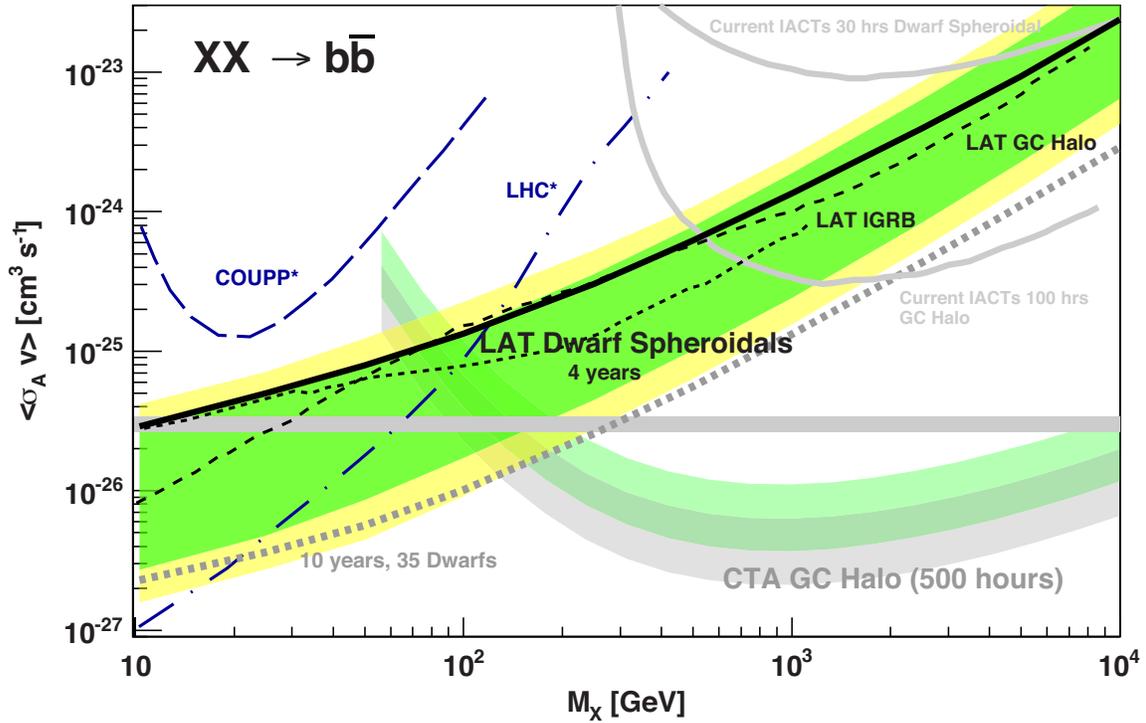}} 
\end{center}
\caption{Compilation of current limits from the Fermi-LAT on dark
  matter annihilation interaction rate $\sigma v$ using gamma rays,
  along with projected future limits from the Cherenkov Telescope
  array (CTA). Also shown are limits from direct detection
  (COUPP~\cite{COUPP}) and accelerator experiments (LHC
  ATLAS~\cite{ATLAS} and CMS~\cite{CMS}) transformed to limits on the annihilation cross
  section on the assumption of four-particle contact interactions as
  in ~\cite{Arrenberg}.  The Fermi-LAT limits were taken
  from the Fermi-LAT paper on dwarf spheroidals~\cite{FermiDwarfs}
  (solid black), on the isotropic diffuse~\cite{FermiIGRB} (dotted),
  on the galactic center halo for an NFW profile~\cite{FermiHalo}
  (dashed).
  The yellow (95\%) and green (68\%) bands give the range of expected
  limits when repeating the procedure multiple times in Monte-Carlo
  simulations on dwarf spheroidals without the inclusion of a signal.
  The light green (Einasto) and light gray (NFW) band give the
  expected CTA sensitivity (500 hours) for the Galactic center halo
  under the assumption of two different dark matter profiles.  For
  these CTA estimates, the contribution by the US groups (doubling the
  number of mid-sized telescopes) was taken into
  account~\cite{CTAUS}. For the Einasto halo model the shape parameter
  was fixed to 0.17. For both models the density was normalized to 0.4
  GeV cm$^{-3}$ at the solar radius. For the sensitivity curves, the
  signal was evaluated by integrating the product of the gamma-ray
  acceptance and the differential DM flux over an annulus of 0.3 - 1.0
  deg from the GC. The background in the signal region was calculated
  from the product of the signal region solid angle and an
  energy-dependent model for the spatial density of protons and
  electrons that survive background rejection cuts.  In addition, the
  uncertainty on the background was modeled by assuming the existence
  of a control region with a solid angle five times that of the signal
  region. }
\label{fig::figure1}
\end{figure*}

\subsection{Targets for dark matter searches}

Because of the quadratic dependence of the self-annihilation rate of
dark matter and thus the gamma-ray flux the detectability of any
particular region in the Universe strongly depends on the density
distribution along the line of sight of the dark matter particles
(so-called J-factor). Unfortunately, dark matter densities are not
very well constrained by numerical simulations, especially in the
innermost regions. In fact, simulations originally showed that the
collapse of cold dark matter gives rise to rather cuspy dark matter
haloes (something that would favor the indirect detection of dark
matter because of the $\rho_\mathrm{DM}^2$ dependency). On the other
hand, observations of galaxy rotation curves favor constant density
cores (so-called `cusp-core problem'', see
e.g. \cite{FloresPrimack1994, Moore1994, NFW1996, Moore1999}).  An
additional complication stems from substructure in the dark matter
distribution that is currently not resolved in cold dark matter N-body
simulations, (i.e. below $\sim10^5 M_{\odot}$). This unresolved
substructure can have a very large impact, in particular in objects
such as galaxy clusters.  Since substructure will further enhance the
annihilation signal this effect is typically quantified in terms of
the so-called boost factor $B$ defined as the ratio of the true
line-of-sight integral to the one obtained when assuming a smooth
component without substructure.  Finally, the situation is further
complicated by the fact that for many objects (such as e.g.\ the Milky
Way) baryonic matter dominates the inner parts of the gravitational
potential. Baryons are expected to have a significant impact on the
dark matter profile compared to the numerical simulations which are
generally dark matter-only. The infall of baryons is expected to alter
the inner dark matter profiles. The profile could either be steepened
through adiabatic contraction~\cite{BaryonsCompressed,
  BaryonsCompressedII, BaryonsCompressedIII}, or it could be flattened
through the occurrence of repeated star bursts triggered by baryonic
infall which tends to render the gravitational potential shallower
since the star burst activity drives out the baryons from the inner
parts~\cite{StarburstCompression, StarburstCompressionII,
  StarburstCompressionIII}.

The choice of the assumed profile of the density distribution
constitutes therefore one of the prime uncertainty in studying dark
matter using gamma rays. As will be discussed later, the uncertainty
on the resulting expected flux limits for individual dwarf spheroidals
are between a factor of 3 for well-constrained objects like Sculptor
up to a factor of 10 for objects such as Coma Berenices. For the
Galactic center, arguably the most promising target in terms of
expected gamma-ray flux from dark matter annihilations, these
uncertainties are considerably larger. See \cite{CatenaUlio2010} for a
discussion on the dark matter profile in the inner Galaxy from a
meta-analysis of kinematic data of the Milky way. Estimates can differ
by up to a factor of 50 depending on the choice of the profile. For
clusters of galaxies the main uncertainties come from the treatment
of substructure below the resolution limit of current numerical
simulations. Also here, the uncertainties in the J-factor can be
several orders of magnitude, and the dark matter profile itself can be
severely modified in these objects by the presence of substructures.
In the following I will go
through indirect detection observations of the various targets,
highlighting current upper limits to WIMP annihilation/mass
cross-sections.

\subsection{Dwarf Spheroidals}

In contrast to the aforementioned uncertainties in the inner parts of
the density profile of objects on Galaxy or galaxy cluster scales,
dwarf spheroidal galaxies can represent a very clean system to search
for dark matter annihilation. Indeed, star formation is usually very
much suppressed in these objects, so astrophysical foregrounds that
produce gamma rays are less of an issue in these objects. From the
stellar kinematics the DM content can appropriately be determined, and
these objects have been found to be the ones with the largest
mass-to-light ratios in the Universe. Boost factors through
substructure below the resolution of numerical simulations are
expected to be irrelevant in these objects and therefore do not add a
large uncertainty. Uncertainties related to the shape of the dark
matter profile are generally integrated over and therefore introduced
by the choice of the profile are at the 10-50\% level. Currently,
there are roughly 25 known dwarf satellite galaxies to the Milky Way
and both ground-based instruments such as H.E.S.S., MAGIC and VERITAS
as well as the Fermi-LAT are actively observing these objects. While
none of the objects are detected with the current generation of
gamma-ray instruments, important conclusions on the properties of dark
matter particles can be drawn from these objects. In particular, a
combined analysis of all known dwarf satellites with the Fermi-LAT
have pushed, for the first time the annihilation interaction rate limits
below the canonical thermal relic production cross-section of $3\times
10^{-26} \mathrm{cm}^3 \mathrm{s}^{-1}$ for a range of WIMP masses
(around 10~GeV) for the annihilation into
$\mathrm{b}\bar{\mathrm{b}}$, which often acts as a
benchmark~\cite{FermiDwarfs, GeringerSamethDwarfs,
  MazziottaDwarfs}. This statement holds also if uncertainties in the
J-factors for these objects are included. Given the all-sky capability
of the Fermi-LAT, a combined analysis of these objects will remain the
cleanest target in the future where more dwarfs are expected to be
detected with future optical surveys such as
Pan-STARRS~\cite{PanStarrs}, DES~\cite{DES} and LSST~\cite{LSST}. It
has been estimated that DES might discovery 19 to 37 new dwarf
galaxies during the duration of the Fermi-LAT
mission~\cite{Tollerud}. At higher energies ground-based Imaging
Atmospheric Cherenkov telescopes (IACTs) have observed dwarf
spheroidals but have not found a signal~\cite{VERITASDwarfs,
  HESSDwarfs, MAGICDwarfs}. Their limits for high WIMP masses are
typically several orders of magnitudes away from the thermal relic
interaction rate and are therefore not (yet) competitive with limits from
the Fermi-LAT at lower energies.

\subsection{Galaxy Clusters}

Galaxy clusters are the largest massive objects in the
Universe. Galaxy clusters are more distant than dwarf spheroidal
galaxies or any of the other targets that are generally used for dark
matter studies using gamma rays. However, like dwarf spheroidals, they
are likely to be dark matter dominated. The range of proposed boost
factors due to unresolved dark matter substructure can be
large. Depending on the assumption about the substructure Galaxy
clusters become competitive in their expected annihilation signal with
dwarf spheroidals only at the extreme (high) end of boost factors. The
best candidate are massive nearby clusters such as Virgo, Fornax or
Coma~\cite{Pinzke11, SanchezConde, GAO11}. One complication for a
possible detection is that galaxy clusters are also expected to
contain a significant number of astrophysical sources of gamma rays,
such as Active Galactic Nuclei (AGN) or radio galaxies. In addition,
these objects are expected to harbor a significant population of
cosmic rays which should radiate gamma rays through interaction with
hadronic material and subsequent pion-decay. As long as no signal is
found, ignoring this contribution represents a conservative assumption
and is therefore justified~\cite{FermiGalaxyClustersI,
  FermiGalaxyClustersII, Huang, Nezri, SanchezConde}. Early claims of
a signal in the Fermi-LAT data from the Virgo cluster~\cite{Han}
turned out to be due to a several unmodeled point-sources within the
cluster~\cite{MaciasRamirez}. At higher energies ground-based
instruments have pushed for rather stringent gamma-ray flux limits on
galaxy clusters (e.g.\ the MAGIC telescopes for the Perseus
cluster~\cite{MAGICPerseus}, the VERITAS array for the Coma
cluster~\cite{VERITASComa}, and the H.E.S.S.\ array for the Fornax
cluster~\cite{HESSFornax}). However, when making conservative
assumptions about boost factors in these objects, the limits on the
benchmark $\mathrm{b}\bar{\mathrm{b}}$ annihilation channel are
several orders of magnitude away from the canonical thermal relic
interaction rate.

\subsection{Isotropic diffuse emission}
The Fermi-LAT has provided a measurement of a faint diffuse isotropic
signal from all over the sky. This so-called isotropic gamma-ray
background (IGRB) follows a featureless powerlaw from 200~MeV to
100~GeV~\cite{FermiEGB} with a photon index of 2.4. This signal is
expected to contain a contribution of mainly extragalactic unresolved
(sub-threshold) sources combined with potentially truly diffuse
emission.  An analysis of the populations of the most numerous sources
detected by Fermi during the first years -- blazars -- showed that
unresolved such objects contribute at most 30\% of the IGRB
emission~\cite{FermiPopulation, FermiPopulationII}.  It is thus
possible that the IGRB emission contains the signature of some of the
most powerful and interesting phenomena in astroparticle
physics. Intergalactic shocks produced by the assembly of Large Scale
Structures~\cite{LoebWaxman2000, Miniati2002, GabiciBlasi2003},
gamma-ray emission from galaxy clusters~\cite{BerringtonDermer2003,
  Pfrommer2008}, emission from starburst and normal
galaxies~\cite{PavlidouFields2002, Thompson2007}, are among the most
likely candidates for the generation of diffuse GeV emission. In
addition, and most relevant for this review, a signal from dark matter
annihilation could be imprinted in the IGRB. While it would be
extremely difficult to detect a dark matter contribution in the IGRB,
upper bounds on dark matter annihilation can be readily derived. The
most conservative approach when calculating upper limits on the dark
matter annihilation interaction rate is to assume that all of the IGRB is
caused by dark matter annihilation. When making rather conservative
assumptions about the contribution of source populations to the IGRB
dark matter annihilation interaction rate limits can be
derived~\cite{FermiIGRB, KevorkIGRB, BringmannIGRB, HooperIGRB} that
are competitive with other methods, such as dwarf spheroidal
galaxies. Obviously, these limits can be significantly tightened when
including additional source populations - however, the degree to which
the contribution from such classes can be determined is questionable.
However, even for the more conservative limits, when combined with
H.E.S.S.\ observational constraints from the Galactic center halo, the
IGRB fluxes rule out all interpretations of the PAMELA positron excess
based on dark matter annihilations into two final lepton
states~\cite{KevorkIGRB} and most of the parameter space for
annihilation into four leptons through new intermediate states.

The statistical properties of the IGRB additionally encodes
information about the origin of this emission. Unresolved sources are
expected to induce a different level of small-scale anisotropies
compared to truly diffuse contributions. A study of the angular power
spectrum of the diffuse emission at Galactic latitudes $|b| >
30^{\circ}$ between 1 and 50 GeV revealed angular power above the
photon noise level at multipoles $l > 155$ independent of
energy~\cite{FermiIGRBAnisotropy}. The scale independence of the
signal suggests that the IGRB originates from one or more unclustered
populations of point sources.  The absence of a strong energy
dependence suggests that a single source class that provides a
constant fractional contribution to the intensity of the IGRB over the
energy range considered may provide the dominant contribution to the
anisotropy. Recently it has been suggested that a strong
correlation between cosmic shear (as measured by galaxy surveys like
DES and Euclid) and the anisotropies in the IGRB might add an
additional handle on the contribution of dark matter annihilation to
the IGRB~\cite{IGRBCosmicShear}.

Because of their limited field of view and the irreducible
electron-shower background, ground based instruments so far have not
provided a competitive measurement of the IGRB. However, the Fermi-LAT
is expected to eventually measure the signal up to several TeV.

\subsection{Galactic Center}
The Galactic center is expected to be the brightest source of dark
matter annihilation gamma rays by at least two orders of
magnitude. However, a multitude of astrophysical sources of gamma rays
in that region complicate the identification of any source region with
dark matter annihilations. For GeV gamma rays the situation is further
complicated by the presence of a highly structured and extremely
bright diffuse gamma-ray background arising from the interaction of
the pool of cosmic rays with dense molecular material in the inner
Galaxy. These astrophysical foregrounds are expected to be several
orders of magnitude brighter in gamma rays than the signal from dark
matter annihilations. Because of these, searches for dark matter
annihilation are usually performed in regions excluding the very
center of the Galaxy. In addition, because of the proximity the
gamma-ray distribution can be expected to be resolved and therefore
the exact choice of the expected radial profile of the dark matter
distribution has a rather large effect when deriving the limits (in
choosing the optimal extraction region).  Data from the Fermi LAT have
been used to search for an annihilation signal from the galactic dark
matter halo~\cite{FermiHalo} and also from the central part of the
Galaxy~\cite{FermiGC}. In both cases rather conservative limits that
assume only the Galactic diffuse emission and dark matter annihilation
to contribute to the observed signals provide limits that are
comparable to those reached by the stacking of dwarf spheroidals. Of
course, these limits can be pushed further down with increasing
assumptions about the objects producing the observed gamma-ray
emission in the galactic center~\cite{HooperLimits}. However,
increasing the complexity of the ``fore-ground modeling'' will also
make these limits less robust. 

At TeV energies, the H.E.S.S.\ telescope system has detected a
point-source coinciding with the supermassive black hole in the center
of our Galaxy~\cite{HESSSgrAStar} and a diffuse
emission coinciding with molecular material in the Galactic
ridge~\cite{HESSDiffuse}. The galactic center source has a featureless
powerlaw spectrum at TeV energies with an exponential cutoff at $\sim
10$ TeV which does not lend itself easily to a dark matter scenario
and is therefore generally thought to be either related to the
supermassive black hole Sgr~A$^{\star}$~\cite{HESSGC} or a pulsar wind
nebula in that region~\cite{HintonAharonian}. Because of the presence
of these bright sources the search for a dark matter signal has
focussed on an angular region of $0.3^{\circ}-1.0^{\circ}$ around the
galactic center. Using 112 hours of observation time, the H.E.S.S.\
collaboration has set the most constraining limits on the annihilation
interaction rate for masses $>1$~TeV, reaching $\sim 7\times10^{−25}
\mathrm{cm}^2 \mathrm{s}^{−1}$ at 1 TeV for the
$\mathrm{b}\bar{\mathrm{b}}$ channel~\cite{HESSHalo}.

Even more interesting, there have been two hints of a signal in the
galactic center region. First, there were reports of an extended
signal coinciding with the center of our
Galaxy~\cite{HooperGoodenough, HooperLinden, Kevork_ManojGC, GordonGC}
above the galactic diffuse emission. There are various alternatives
for the origin of this signal, amongst others the interaction of
freshly-produced cosmic rays with interstellar
material~\cite{HooperGoodenough, Kevork_ManojGC, LindenLovegrove,
  LindenProfumo}, a population of mili-second
pulsars~\cite{KevorkMSPulsars, WhartonMSPulsars, JennyMSPulsars}, or
the annihilation of dark matter particles with masses between 7 and 40
MeV~\cite{HooperGoodenough, HooperLinden, Kevork_ManojGC}. While the
signal appears to be real and naturally is of great interest, the lack
of a smoking gun feature that could help relate it to dark matter
annihilation will make it extremely difficult to convincingly and
unambiguously claim a dark matter detection unless backed by other
measurements.

In that regard, the second claimed signal in the galactic center
region is even more exciting, the discovery of a line signal at $\sim
130$ GeV in an extended region around the galactic center. I will
describe its general properties in the following section.

\subsection{Line Searches}

The annihilation of dark matter into $\gamma$X leads to monochromatic
gamma rays with $E_\gamma = m_\chi(1-m_X/4 M_\chi^2)$. Such a signal
will provide a smoking-gun signal that is very difficult to mimic in
astrophysical sources, in particular if found in more than one
location on the sky. This process is expected to be strongly
loop-suppressed by a factor $\mathcal{O}(\alpha^2_{\mathrm{e}})$. The
discovery of a hint of a signal at $\sim 130$ GeV in the Fermi-LAT
data in a region-of-interest (ROI) optimized for particular dark
matter distributions towards the Galactic center~\cite{BringmannJCAP,
  WenigerJCAP} have raised the exciting possibility that if confirmed
this could be the long awaited first clear evidence of dark matter
annihilation into gamma rays. While very serious doubts about the
astrophysical origin of the signal have been raised recently as will
be discussed below, the statistical significance of the original
signal seems to be beyond the level of a statistical fluctuation. Two
obvious alternatives to a dark matter annihilation signal exist: (1)
an astrophysical origin (such as e.g.\ inverse Compton scattering in
the Klein-Nishina regime of mono-energetic electrons produced in
pulsar winds~\cite{AharonianKangulyan}) and (2) one or more systematic
(i.e.\ instrumental) effects. Given that the signal was found to be
significantly extended, the pulsar hypothesis can safely be discarded
and therefore the only viable alternative explanation is an
instrumental effect.

Although only marginally significant (claimed post-trial significance
of 3.2$\sigma$), the community has enthusiastically responded to this
signal. The extension of the signal seems to be compatible with
conventional dark matter profiles such as NFW or Einasto and slightly
offset but generally in the direction of the Galactic
center~\cite{SuFinkbeiner, RaoWhiteson} -- although claims were made~\cite{Yang}
that this offset could be due to limited statistics of the signal
($\sim 14$ gamma-ray photons). There was a mild tension with
all-sky limits for dark matter annihilation lines released around the
same time by the Fermi-LAT team~\cite{FermiLineSignal} but that could
be argued was due to the difference in extraction region (all-sky
versus optimized ROI for a particular dark matter profile). Additional
studies quickly suggested a second line at 114 GeV that -- in the
picture of dark matter annihilation -- was identified as the $\chi\chi
\leftarrow \gamma$Z$^0$ line~\cite{SuFinkbeiner} (although only at the
$1-2 \sigma$ level). Line signals at 130 GeV at low significance were
also claimed to be detected from unidentified
sources~\cite{FinkbeinerUnid} and from galaxy
clusters~\cite{HektorRaidalTempel2012}. 

Compared to these early indications, the picture has, however, become
significantly murkier at closer inspection -- instrumental effects do
seem at least to play a role in this signal. Early doubts that the
signal could be a contamination of the background estimate arising
from the spectral shape of the Fermi bubbles (power law with sharp
break) in that region~\cite{ProfumoBubblesLine} were likely ruled out
based on the morphology of the signal~\cite{Tempel2012}. The spectrum
of most of the unidentified sources that showed the weak hint of two
lines at similar energies to the original galactic center signal were
shown to be incompatible with a dark matter annihilation
origin~\cite{HooperLinden, Mirabal} but rather in agreement with those
from active galactic nuclei (AGN). This finding therefore challenges
the notion of the signal arising from DM annihilations and makes an
instrumental effect more likely. Also, the galactic center source
began at least partially to unravel -- the growth of the signal over
time after the first discovery did not follow the expected trend but
seemed rather compatible with a background fluctuation (see e.g.\
~\cite{WenigerFermiSymposium}). 

A similarly strong challenge to the dark matter interpretation came
from an updated analysis by the Fermi-LAT
collaboration~\cite{FermiLinesUpdate}. First, when taking the energy
dispersion of the detector into account in the fitting, the signal
significance decreases. Essentially, the signal is too narrow to be a
genuine signal. In fact, a search in optimized ROIs over the whole
energy range by the Fermi-LAT collaboration, taking into account the
energy dispersion in the fitting, does not detect any significant lines
at any energy above the $2\sigma$ level when accounting for all the
trials factors~\cite{FermiLinesUpdate}.  Similarly important, when
analyzing the Earth limb data -- arising from the interaction of
cosmic rays with the atmosphere -- as a reference data set free of
dark matter interactions shows a weak ($\sim 3\sigma$) line signal at
130 GeV for a certain range of incidence angles ($<60^{\circ}$) in the
detector~\cite{Finkbeiner}. Given that the expected Earth limb signal
is a featureless powerlaw in that range~\cite{FermiEarth} such a
feature points towards an instrumental effect that preferentially
reconstructs events at those energies. However, this (currently
unknown) effect can only explain a minor fraction ($\sim 20\%$) of the
galactic center signal. A second reference region (the
galactic disk excluding the galactic center region) does not show such
a feature at 130 GeV or at any other energy.

Given these complications, the current situation is such that there
are some serious doubts about the origin of the signal, however,
instrumental effects can also not fully explain the occurrence of the
signal and therefore a dark matter origin remains a valid
possibility. Future Fermi-LAT data will clarify the situation. A
completely rewritten event reconstruction of the Fermi-LAT data that
remedies several problems found after launch will enhance the energy
resolution and will become available later this year (so-called
\emph{Pass-8}). If the signal remains significant with these updates,
the Fermi-LAT management is considering an updated observing strategy
that would enhance the exposure to the Galactic center. If the signal
is not too extended ($\lessapprox 2^{\circ}$ FWHM), the HESS-II array
with its new 27-m diameter dish will have the required sensitivity to
independently rule out or confirm the line at
130~GeV~\cite{Bergstroem2012}.

\section{Future Searches and complementarity with other methods}
\label{sec::future}

In the absence of a space-mission that would improve the overall
sensitivity over the Fermi-LAT significantly in the near future the
community is looking toward the next generation ground-based
instrument as the next big step in the indirect detection of dark
matter through gamma rays\footnote{it should be noted, that the
  planned Russian/Italian space-mission
  {\emph{Gamma-400}}~\cite{Gamma400} aims to significantly improve the
  energy resolution which might be relevant for line searches}. The
Cherenkov Telescope Array (CTA) is expected to start operation later
in this decade (current start of construction planned for 2016) and
will have sensitivity over the energy range from a few tens of GeV to
100s of TeV. To achieve the optimal sensitivity over that wide a
range in energy, CTA will employ three different telescope sizes: Large
Size Telescope (LST, 23 m diameter), Medium Size Telescope (MST, 10-12
m) and Small Size Telescope (SST, 4-6 m). The design goal is a
point-source sensitivity of at least an order of magnitude better than
currently operating instrument at the sweet-spot of 1 TeV and a
significantly improved angular resolution, improving with energy from
0.1$^\circ$ at 100 GeV to better than 0.03$^\circ$ at energies above 1
TeV. The US groups within the CTA consortium are planning to augment
the array in the crucial mid-size telescope part. Current plans call
for a doubling of the baseline number of mid-sized telescopes,
enhancing the sensitivity by a factor of 2--3~\cite{CTAUS}.

Gamma rays are sensitive to almost any annihilation channel with a
sensitivity that is closely related to the total annihilation cross
section of dark matter that underlies its total relic abundance
today. Already now, gamma-ray limits are probing below the thermal
relic interaction rate for some of the preferred WIMP mass range. While
direct detection experiments mostly have to deal with uncertainties in
the background estimations, the indirect detection technique is
largely dominated by astrophysical uncertainties. For dwarf spheroidal
galaxies, these are largely mitigated by the lack of astrophysical
backgrounds and tight constrains on the halo profile from dynamical
measurements. For the galactic center these uncertainties are the
largest but the prospects for a detection are still the highest. A
positive (and credible) detection would entail either a gamma-ray
line, like in the case of the aforementioned one at 130~GeV in the
Fermi-LAT data or alternatively, from the measurement of identical
spectra that are compatible with a dark matter origin from more than
one source. None of the proposed dark matter search methods (direct
detection, indirect detection or accelerator searches) will be able to
unambiguously claim the detection of dark matter and thus all the
methods are crucial in a viable dark matter program for the
future. Each potential signal will potentially be created by a new
(previously unknown) background -- even in the case of accelerator
searches. One big advantage of the indirect detection techniques is
that if a signal is found in an accelerator or in a direct detector,
gamma-ray measurements will provide the only way to connect the
laboratory to the actual distribution of dark matter on the sky and
identify the nature of the particle through the details of the
annihilation process. In fact detecting a signal from the galactic
center would allow to measure the dark matter density profile and feed back to
cosmological simulations.

\begin{figure*}[t]
\begin{center}
\resizebox{2 \columnwidth}{!}{\includegraphics{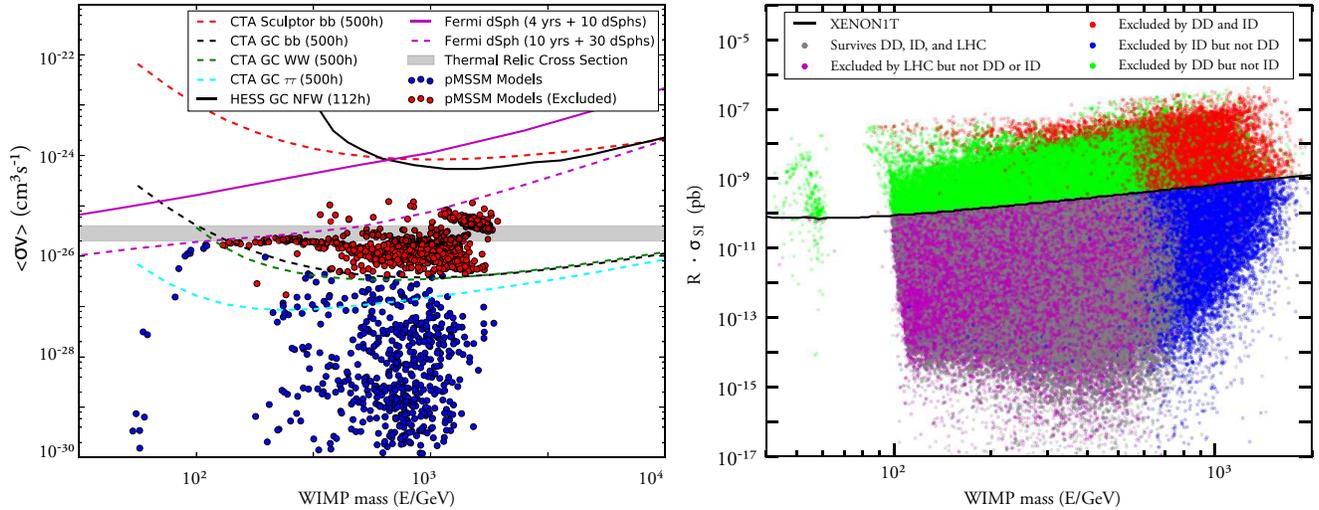}}
\end{center}
\caption{{\bf{Left:}} Reproduced from~\cite{SnowmassCTA}. Comparison
  of current (solid lines) and projected (dashed lines) limits on the
  DM annihilation interaction rate from different gamma-ray searches as a
  function of WIMP mass.  Limits for Fermi (magenta lines) and
  H.E.S.S. (solid black line) are calculated for a 100\% branching
  ratio to $bb$ (same as in Figure 1). Projected limits for CTA are
  shown for WIMP annihilation to $bb$ and a 500 hour observation of
  Sculptor (red dashed line) and for WIMP annihilation to $bb$ (black
  dashed line), $W^{+}W^{-}$ (green dashed line), and
  $\tau^{+}\tau^{-}$ (cyan dashed line) and a 500 hour observation of
  the GC. For the sensitivity calculation of CTA the baseline array
  I~\cite{CTADesignStudy} supplemented by a contribution of 36
  mid-sized dual-mirror telescopes by the US groups is assumed.  The
  calculation of the annihilation flux for the GC region assumes an
  NFW MW halo profile with a scale radius of 20 kpc and DM density at
  the solar radius of 0.4 GeV~cm$^{-3}$. Filled circles represent
  pMSSM models satisfying WMAP7 constraints on the relic DM density
  and experimental constraints from ATLAS and CMS SUSY searches and
  XENON100 limits on the spin-independent WIMP-nucleon cross section
  \cite{2011EPJC...71.1697C,2012EPJC...72.2156C}.  Models indicated in
  red would be excluded by the CTA 95\% C.L. upper limit from a 500
  hour observation of the Galactic Center.  {\bf{Right:}} Reproduced
  from ~\cite{CompSnowmass}. Comparison of exclusion ranges for the
  three dark matter search strategies: direct detection (XENON1T and
  COUPP500 in green and red), indirect detection (CTA and Fermi-LAT in
  red and blue) and collider searches (LHC up to 8 TeV in
  magenta). Each dot represents a currently allowed pMSSM model in the
  plane of spin-independent cross-section versus mass of the lightest
  supersymmetric particle. Gray dots will not be detectable by any
  method.}
\label{fig::figure2}
\end{figure*}

In addition, there is a unique region of the WIMP parameter space that
CTA can best address in the near future -- the high-mass ($\sim 1$ TeV)
scenario. Figure~\ref{fig::figure2} (reproduced
from~\cite{SnowmassCTA} and ~\cite{CompSnowmass}) shows the reach of
CTA in currently-allowed pMSSM models as a function of the WIMP
mass. The pMSSM (for \emph{phenomenological} MSSM) is a 19-parameter
model~\cite{Djouadi1998} that represents the most general version of
an R-parity conserving MSSM subjected to experimental constraints. No
assumptions are made about the physics of SUSY breaking within the
pMSSM models. Therefore the pMSSM model set shows a much broader range
of phenomenology than found in highly-constrained model sets. The
left-hand graph shows the CTA sensitivity in the traditional
representation in the $<\sigma_{\mathrm{ann}} v>$ WIMP mass plane (similar to
Figure~\ref{fig::figure1}) for the $b\bar{b}$ annihilation channel
overlaid with pMSSM models that saturate the thermal relic density
(i.e.\ these particle constitute the bulk of the cosmological dark
matter). The right-hand graph shows the same sensitivity, this time in
the plane that is traditionally associated with direct-detection
experiments, the spin-independent scattering cross-section
$\sigma_{\mathrm{SI}}$ for pMSSM models without a constraint on the
thermal relic density. As can be seen, CTA has significant reach in
this model parameter space and is in particular uniquely suited to
sensitively address WIMP masses around 1~TeV and above.

\section{Conclusion}
This is a exciting time for the search for dark matter. As the title
of the NAS symposium suggests: we might be on the threshold of a
discovery because our experimental and observational capabilities have
progressed to the point where all three legs of the search for dark
matter are sensitive to testing the WIMP paradigm in a very serious
way. The LHC detectors are used to search for super-symmetric
particles and will dominate the accelerator-based search for the next
decade. The direct detection experiments are moving to ton-scale
detectors and have taken great strides in improving their
understanding of the backgrounds. And finally, indirect detection
experiments, in particular using gamma rays, such as the Fermi-LAT,
H.E.S.S., MAGIC, and VERITAS and the future Cherenkov Telescope Array
(CTA) are starting to probe the thermal relic interaction rate in various
astrophysical objects. While uncertainties about the dark matter
profiles and the astrophysical foregrounds will always have a serious
impact on the detectability of dark matter annihilation signals, the
limits have gotten robust against these problems most recently. If
lucky, we might even be able to measure the dark matter density
profiles in the very inner parts of our galaxy or in dwarf satellites
using gamma rays.






\begin{acknowledgments}
  I would like to thank Roger Blandford, Rocky Kolb, Maria Spiropulu,
  Michael Turner, and the National Academy of Sciences for the
  opportunity to present this summary at the NAS Sackler Symposium on
  Dark Matter in Irvine, California. I am grateful for discussions on
  the subject with Elliott Bloom, Alex Drlica-Wagner, Seth Digel,
  Michael Peskin, Miguel S\'{a}nchez-Conde, Justin Vandenbroucke, Risa
  Wechsler, and Matthew Wood. I would like to thank Miguel
  S\'{a}nchez-Conde, and Michael Peskin for their reading of the
  manuscript and their helpful comments.
\end{acknowledgments}





\end{article}
\end{document}